%% file: latex/acl_latex.tex
\pdfoutput=1

\documentclass[11pt]{article}
\usepackage{booktabs}
\usepackage{amsmath}
\usepackage{amsthm}
\usepackage{algorithm}
\usepackage{algpseudocode}

\newtheorem{definition}{Definition}
\usepackage{multirow}
\usepackage{graphicx}
\usepackage{subcaption}

\usepackage[preprint]{acl}

\usepackage{times}
\usepackage{latexsym}

\usepackage[T1]{fontenc}

\usepackage[utf8]{inputenc}

\usepackage{microtype}

\usepackage{inconsolata}

\usepackage{graphicx}

\usepackage{enumitem}

\title{Ensemble Privacy Defense for Knowledge-Intensive LLMs against Membership Inference Attacks}

\author{
    Haowei Fu\textsuperscript{1}\thanks{Equal contribution.} \quad
    Bo Ni\textsuperscript{1}\footnotemark[1] \quad
    Han Xu\textsuperscript{2} \quad
    Kunpeng Liu\textsuperscript{3} \quad
    Dan Lin\textsuperscript{1} \quad
    Tyler Derr\textsuperscript{1} \\[0.3em]
    \textsuperscript{1}Vanderbilt University \quad
    \textsuperscript{2}University of Arizona \quad
    \textsuperscript{3}Clemson University \\[0.3em]
    \texttt{\{haowei.fu, bo.ni, dan.lin, tyler.derr\}@vanderbilt.edu}\\
    \texttt{hanxu@arizona.edu, kungpeng@clemson.edu}
}
\begin{document}
\maketitle
\begin{abstract}
\input{latex/Sections/0abstract}
\end{abstract}

\section{Introduction}
\input{latex/Sections/1introduction}

\section{Preliminaries}\label{sec:preliminaries}
\input{latex/Sections/2preliminaries}

\section{Robustness of RAG vs. SFT under Membership Inference Attacks}\label{sec:rag_vs_sft}
\input{latex/Sections/3robustness}

\section{Ensemble Privacy Defense} \label{sec:epd_framework} %  for MIAs}

\input{latex/Sections/4epd}

\section{Experiments}\label{sec:experiments} % and Results}
\input{latex/Sections/5experiments}

\section{Related Work}\label{sec:relatedwork}
\input{latex/Sections/relatedwork}

\section{Conclusion}\label{sec:conclusion}
\input{latex/Sections/6conclusion}

%\newpage
\section{Limitations}

While our proposed framework demonstrates significant improvements in privacy protection, several limitations should be acknowledged. First, our evaluation is primarily conducted on RTX4090 GPU machine, so we do not include the baseline retraining model/recreating the dataset, nor model components with greater capacity (e.g., DeepSeek-R1-33B). Second, the computational overhead of the ensemble approach, including the judge model inference, may limit its %practical 
deployment in resource-constrained environments. Third, our noise injection experiments show mixed results, indicating that the effectiveness of additional privacy mechanisms may be context-dependent and require further optimization. Finally, the framework's performance is evaluated against existing MIA attack baselines, but future attacks may exploit different vulnerabilities that our current defense mechanisms do not address.

\section{Ethical Considerations}
 While our research aims to enhance privacy protection, we acknowledge ethical considerations. Our evaluation is limited to specific datasets and model architectures, and the effectiveness of our methods may vary in different contexts. We encourage the research community to consider the broader implications of privacy attacks and defenses in real-world applications, and we commit to responsible disclosure of any significant vulnerabilities discovered during our research.

\bibliography{latex/acl_latex}

\appendix

\input{latex/Sections/appendix}

\end{document}

%% file: latex/Sections/0abstract.tex
Retrieval-Augmented Generation (RAG) and Supervised Finetuning (SFT) have become the predominant paradigms for equipping Large Language Models (LLMs) with external knowledge for diverse, knowledge-intensive tasks. However, while such knowledge injection improves performance, it also exposes new attack surfaces.  Membership Inference Attacks (MIAs), which aim to determine whether a given data sample was included in a model's training set, pose serious threats to privacy and trust in sensitive domains. To this end, we first systematically evaluate the vulnerability of RAG- and SFT-based LLMs to various MIAs. Then, to address the privacy risk, we further introduce a novel, model-agnostic defense framework, Ensemble Privacy Defense (EPD), which aggregates and evaluates the outputs of a knowledge-injected LLM, a base LLM, and a dedicated judge model to enhance resistance against MIAs. Comprehensive experiments show that, on average, EPD reduces MIA success by up to 27.8\% for SFT and 526.3\% for RAG compared to inference-time baseline, while maintaining answer quality. Our code will be made available public at \url{https://github.com/RageFu2004/Ensemble-Privacy-Defense}. 

%% file: latex/Sections/1introduction.tex
Large language models (LLMs) have become the foundation of modern natural language processing, powering applications across diverse domains~\cite{vaswani2017attention, devlin2019bert, brown2020language}. For knowledge-intensive tasks, two predominant paradigms inject external knowledge into LLMs: retrieval-augmented generation (RAG) and supervised fine-tuning (SFT)~\cite{ovadia2024finetuningretrievalcomparingknowledge}. RAG retrieves evidence at inference time and conditions generation on the retrieved passages~\cite{lewis2020retrieval}, whereas SFT adapts a pretrained model to a downstream task using task-specific data~\cite{ouyang2022training}. Both approaches substantially improve factuality, relevance, and adaptability.
\begin{figure}[t!]
\centering
\includegraphics[width=0.9\linewidth]{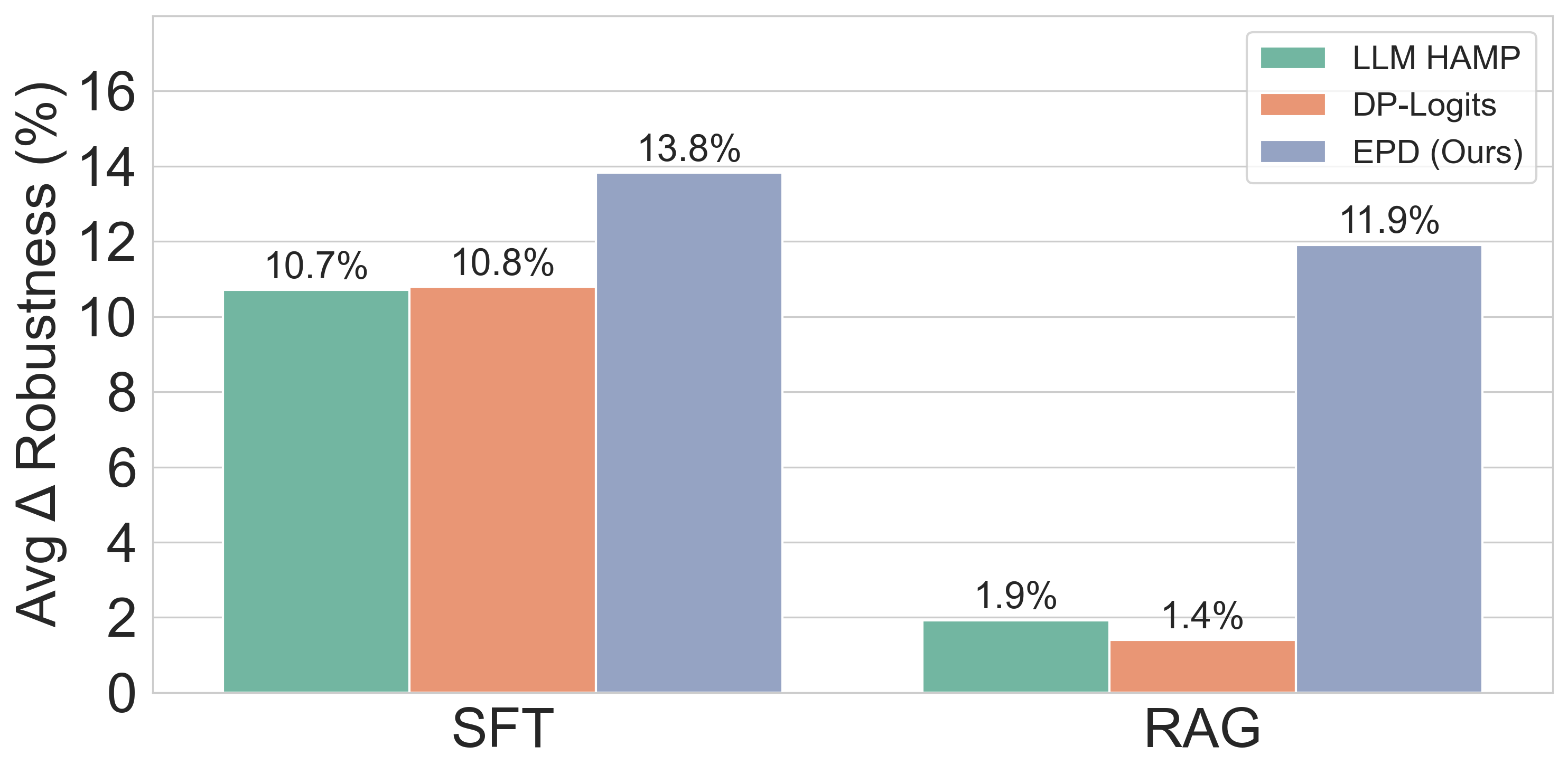}
\vspace{-1ex}
\caption{Comparison of improved robustness on MIAs across datasets. Our proposed method can significantly reduce the risk of privacy leakage under MIAs.}
\vspace{-1ex}
\label{fig:teaser}
\end{figure}
As knowledge-intensive LLMs move into high-stakes settings, however, privacy risks have become paramount~\cite{ni2025trustworthyretrievalaugmentedgeneration, hu2022membership}. A key threat is membership inference attacks (MIAs), which aim to determine whether a particular example was used to train or condition the model~\cite{yeom2018privacy, carlini2021extracting}. RAG and SFT create different attack surfaces~\cite{Anderson_2025}: SFT updates parameters, potentially tightening the coupling between specific fine-tuning data and model outputs; RAG introduces a pipeline to the external corpus where the retrieved contents can leak through generation. In practice, RAG often appears more MIA-resistant because knowledge resides primarily in a non-parametric index, retrieval adds stochasticity, and conditioning on heterogeneous evidence raises output entropy—reducing the separability between member and non-member scores that many MIAs exploit. 

Prior work has studied MIAs and defenses for either RAG~\cite{Anderson_2025} or SFT~\cite{fu2023practical, huang2025dfmia}, but a systematic, head-to-head comparison is lacking, and many defenses are tightly coupled to specific architectures or training objectives~\cite{Anderson_2025, liu2025maskbasedmembershipinferenceattacks, zhang2025softselectivedataobfuscation}, limiting real-world adoption. We address both gaps by (i) providing a controlled comparison of MIA vulnerability across RAG and SFT, and (ii) introducing a training-free, model-agnostic Ensemble Privacy Defense (EPD). Given a query, EPD obtains candidate answers from a knowledge-injected target model and a base model, and uses a judge LLM to select/synthesize the final output. Intuitively, aggregating heterogeneous candidates---and instructing the judge to penalize verbatim phrasing and down-weight abnormally low per-token loss---acts as an entropy-increasing regularizer: it narrows member–non-member likelihood gaps (e.g., LiRA~\cite{carlini2022membership}) 
and weakens tail-token cues (e.g., Min-K~\cite{shi2024min}), 
reducing MIA effectiveness. 

We study two questions: which paradigm—RAG or SFT—is more robust to MIAs, and can a training-free, model-agnostic ensemble mitigate MIAs while preserving answer quality? To answer them, we evaluate RAG and SFT models on multiple QA datasets against seven representative MIAs, and then comprehensively assess EPD defense capability against MIAs. Empirically, RAG generally shows stronger resistance than SFT (with higher latency due to retrieval), and EPD further improves defense abilities in both RAG and SFT. Figure~\ref{fig:teaser} demonstrates that, on average, EPD can successfully improve the robustness of the knowledge-injected models on inference-time, with a 27\% improvement for SFT and 526\% improvement for RAG. These results reveal an efficiency–privacy trade-off and demonstrate that judgment-guided ensembling of EPD offers a practical path to privacy-preserving deployment.
Our contribution can thus be summarized as follows:

\vspace{-0.5ex}
\begin{itemize}[leftmargin=*]
    \item We systematically study MIAs on knowledge-intensive LLMs, analyzing the vulnerabilities of both RAG and SFT paradigms.
    \item Ensemble Privacy Defense (EPD): a training-free, model-agnostic framework that aggregates a target and base model with a judge LLM to attenuate membership signals while preserving utility.
    \item Experiments on QA benchmarks against seven MIAs, demonstrating substantial defense capabilities and providing insights for privacy-preserving RAG-/SFT-based LLM deployment.
\end{itemize}

%% file: latex/Sections/2preliminaries.tex
\subsection{Knowledge-Intensive LLMs} 

LLMs have demonstrated remarkable capabilities across NLP tasks, from text generation and summarization to dialogue systems~\cite{vaswani2017attention, devlin2019bert, brown2020language}. However, despite their impressive performance, traditional LLMs rely solely on their parametric memory, which is limited by their training data cutoff date and the inherent constraints of model capacity, leading to deteriorated performance in domain-specific, knowledge-intensive tasks~\cite{ni2025trustworthyretrievalaugmentedgeneration, lewis2020retrieval, hu2022lora}. To address these limitations, two predominant paradigms have emerged for equipping LLMs with external knowledge: Retrieval-Augmented Generation (RAG)~\cite{lewis2020retrieval} and Supervised Finetuning (SFT)~\cite{ovadia2023fine}. These approaches enable LLMs to access external knowledge sources dynamically, thereby enhancing their factual accuracy, reducing hallucination, and improving their applicability to knowledge-intensive tasks.

\subsection{Membership Inference Attacks}
Here we formally introduce the definition and threat model of MIA on LLMs. 

\begin{definition}[Membership Inference Attacks]
Let $\mathcal{D}_{\text{train}}$ denote the training dataset used to train a large language model (LLM) $f_\theta$, where $\theta$ represents the model parameters. Given a query $q$, the LLM generates a response $a = f_\theta(q)$. Membership inference attacks (MIAs) aim to determine whether a specific data sample $x$ was included in $\mathcal{D}_{\text{train}}$, i.e., to infer the membership status $m(x) \in \{0, 1\}$. Formally, an adversary is given black-box access to the LLM and, for a target sample $x$, attempts to construct an attack function $\mathcal{A}$ such that:
\begin{equation}
    \mathcal{A}(f_\theta, x) \approx m(x) \nonumber 
\end{equation}
\end{definition}
In the context of SFT, $\mathcal{D}_{\text{train}} = \mathcal{D}_{\text{train}}^{\text{SFT}}$, where the training data consists of labeled examples used to adapt a pre-trained model to a downstream task. In the context of RAG, $\mathcal{D}_{\text{train}}$ is replaced by the retrieval corpus $\mathcal{C}$. MIAs here attempt to determine whether a specific document or passage $x$ is contained in $\mathcal{C}$. We assume that the attacker does not have any knowledge of the retrieval model.   

\paragraph{Threat Model.}
We consider a setting where the victim is the large language model $f_\theta$ trained either by SFT on a dataset $\mathcal{D}_{\text{train}}^{\text{SFT}}$ or deployed with RAG using an external knowledge corpus $\mathcal{C}$, and the attacker is an external adversary aiming to determine whether a specific sample $x$ was used in finetuning or whether a document/passage $x$ exists in the retrieval corpus. We assume a black-box setting where the adversary can only query the target model and observe its outputs (e.g., probabilities, likelihoods, or generated text) without access to parameters or training data, but may possess limited auxiliary knowledge such as samples from a similar distribution, domain expertise, or the ability to construct probing queries. 

\subsubsection{MIA Methods} %Attack Methods}
\label{sec:mia_attack_methods}

The MIAs %attacks 
can be categorized into two categories: Reference-free attacks and Reference-based attacks, where Reference-free attacks rely solely on the outputs of the target model, whereas reference-based attacks additionally leverage auxiliary datasets or reference models for calibration. We explore the following common MIA 
methods in the rest of the paper. 

\paragraph{Reference-free attacks.} 
 Representative methods include \textit{Recall}, which leverages the model's recall or confidence on the queried sample~\cite{xie2024recall}; 
\textit{LL (Log-Loss)}, which uses the negative log-likelihood of the model's prediction as a membership signal~\cite{yeom2018privacy}; 
\textit{Zlib}, which measures the compressibility of the model's output since memorized samples tend to be more compressible~\cite{carlini2021extracting}; 
and \textit{Min-K/Min-K++}, which focus on the lowest $K$ token losses or their z-score normalization to detect membership signals~\cite{shi2024min, zhang2024min}.

\paragraph{Reference-based attacks.} 
Representative methods include 
\textit{SPV-MIA (Self-calibrated Probabilistic Variation)},  which constructs a reference dataset by prompting the target LLM itself and introduces a probabilistic variation metric for more reliable membership signals in practical scenarios~\cite{fu2023practical}; 
\textit{LiRA}~\cite{carlini2022membership} compares the output of the target model to that of a reference model trained on a different dataset.

%% file: latex/Sections/3robustness.tex
To evaluate the robustness of RAG and SFT against membership inference attacks, we benchmark them on a set of common MIAs. We first detail the experimental setup and datasets, then present results and analysis. For fairness, all methods are evaluated under an identical protocol with matched hyperparameters, prompts, and inference budgets; the same preprocessing, attack implementations, and metrics are used unless noted. This controlled design enables a direct comparison of RAG and SFT robustness across a range of MIA variants.
\begin{figure}[t]
    \centering 

    \fbox{
        \tiny
        {\color{blue!60}\rule[0.5ex]{1em}{0.5pt}} SFT \quad 
        {\color{orange}\rule[0.5ex]{1em}{0.5pt}} RAG
    }

    \begin{subfigure}[t]{0.48\columnwidth}
        \centering
        \includegraphics[width=\linewidth]{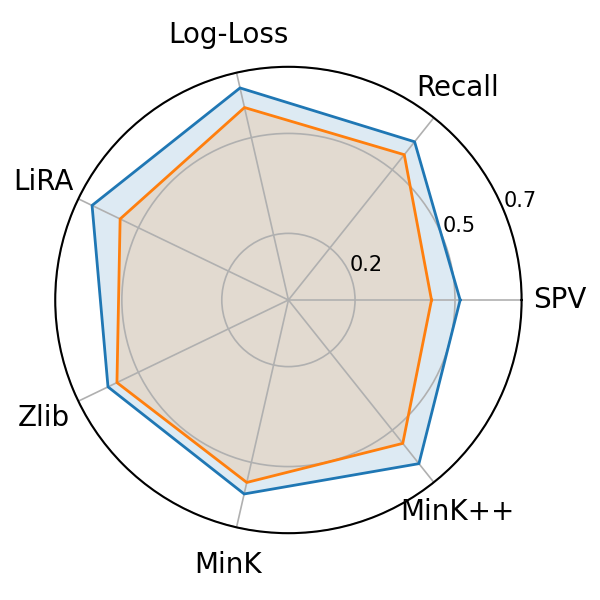}
        \caption{Attack Success Rate}
        \label{fig:mia_sft}
    \end{subfigure}
    \hfill
    % Second subfigure
    \begin{subfigure}[t]{0.48\columnwidth}
        \centering
        \includegraphics[width=\linewidth]{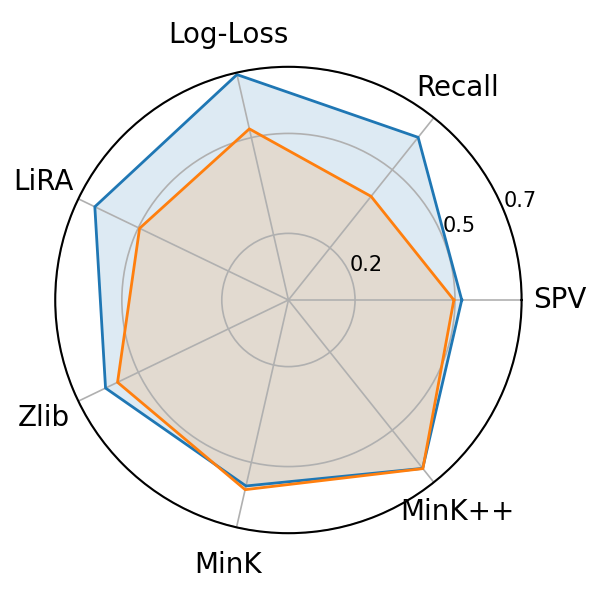}
        \caption{AUC Score}
        \label{fig:mia_rag}
    \end{subfigure}

    \caption{
    Robustness of RAG and SFT, averaged over three datasets, against seven MIAs, shown in terms of (A) attack success rate and (B) AUC score.}
    \label{fig:mia_sft_rag}
    \vspace{-1.1em}
\end{figure}
\subsection{Findings}

\subsection{Knowledge-Intensive LLM Settings}
\label{robustness:implementation}
\paragraph{Supervised Finetuning.} We employ Llama-2-7B~\cite{touvron2023llama2openfoundation} as the base model for supervised finetuning. For model finetuning, we utilize a Parameter-Efficient Fine-Tuning (PEFT) technique, specifically LoRA~\cite{hu2022lora}, to update the model parameters. 

\paragraph{Retrieval Augmented Generation.}
For fair comparison, we use the Llama-2-7B as the LLM for inference. For dense retrieval, we employ the SentenceTransformer~\cite{reimers2019sentencebertsentenceembeddingsusing} to embed both the input queries and all context passages from the datasets. At inference time, we compute the cosine similarity between the query embedding and all stored context embeddings, retrieving the top-$K$ most similar contexts (with $K=5$) to construct the final prompt for the LLM. 

\subsection{Datasets and Metrics}
We utilize two widely used datasets for knowledge-intensive LLMs~\cite{ovadia2023fine, kim2024reragimprovingopendomainqa}, 
more specifically, TriviaQA~\cite{joshi2017triviaqalargescaledistantly} and SQuAD~\cite{rajpurkar2016squad100000questionsmachine}. Both question answering datasets have a database constructed from Wikipedia to help answer the queries. The dataset statistics are presented in Table~\ref{tab:datasets}, and more details are provided in Appendix~\ref{sec:dataset_details}.

For evaluation, we leverage the commonly used metrics for privacy attacks: AUC and ASR. Area Under Curve (AUC) measures the overall discriminative ability of the MIA attacks by calculating the area under the ROC curve, i.e., the probability that a randomly chosen member receives a higher attack score than a non-member, and Attack Success Rate (ASR) is the proportion of successful membership inference attempts.

We report the average ASR and AUC across the two datasets in Figure~\ref{fig:mia_sft_rag}, with the full detailed results supplemented in Appendix Table~\ref{tab:baseline_triviaqa_squad_comparison}. Our findings reveal that RAG demonstrates consistently stronger resistance to MIA attacks compared to SFT, particularly in terms of lowering ASR. On Trivia-QA, RAG achieves substantially lower ASR values across nearly all attack methods while maintaining lower overall AUC scores. On SQuAD, RAG also shows advantages under most attack settings. Averaged across both datasets, RAG can reduce ASR and AUC by nearly 10\%, demonstrating its robustness as a more privacy-preserving paradigm. The unfinetuned LLM in RAG means that the context are supplemented explicitly, without changing its parameters as in SFT. This helps avoid member data memorization while the retrieved contexts can provide same, or even more relevant information for the downstream generation.  

\begin{figure}[t!]
\centering
\includegraphics[width=0.97\linewidth]{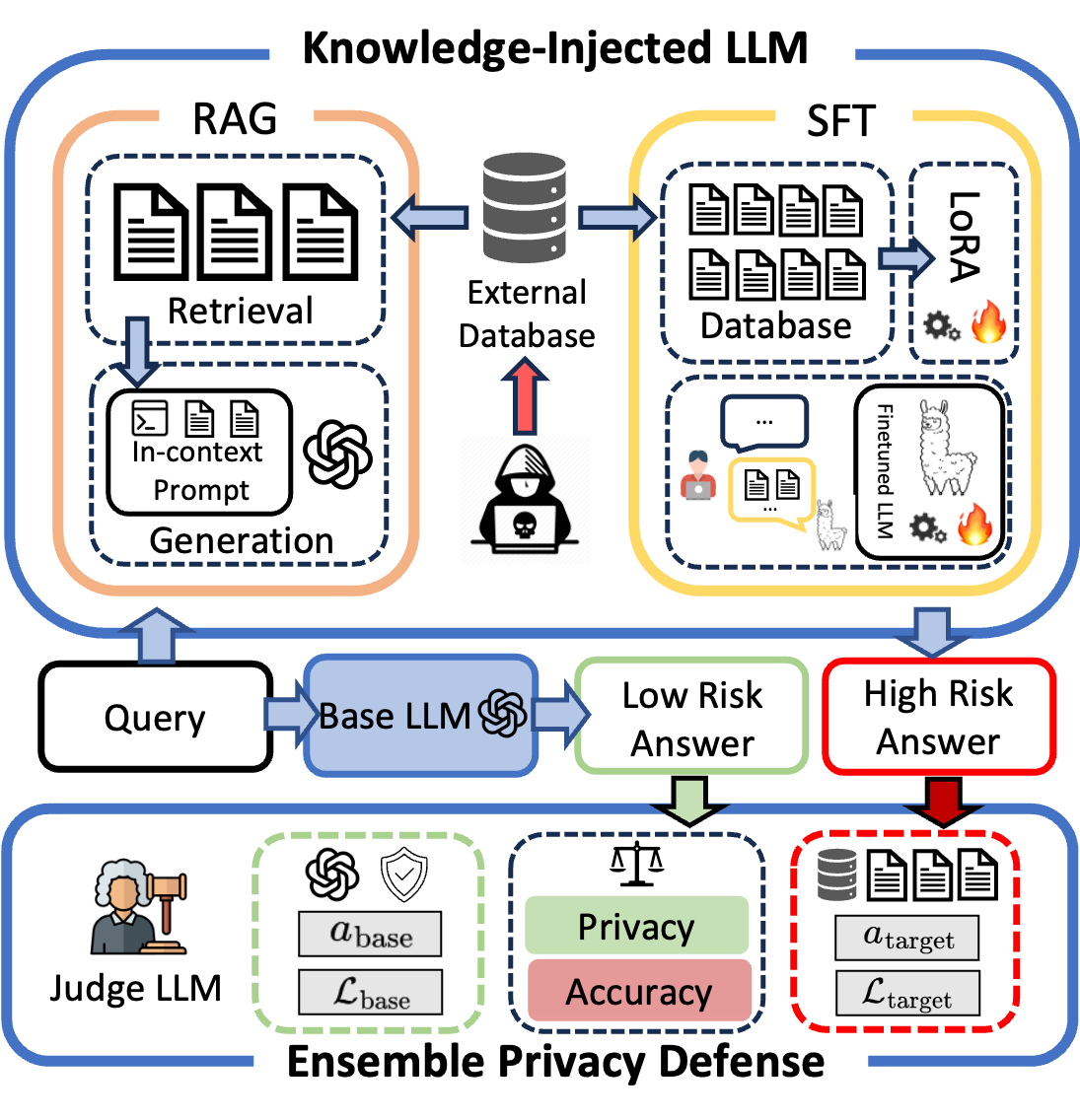}
\vspace{-1.25ex}
\caption{Overview of the proposed EPD framework.}
\vspace{-1.25ex}
\label{fig:framework}
\end{figure}

Nevertheless, knowledge-intensive LLMs remain exposed, with ASR consistently exceeds 50\% across both RAG and SFT settings. Existing defense strategies are often tightly coupled with model architectures and require additional training or fine-tuning on the base model~\cite{Anderson_2025, liu2025maskbasedmembershipinferenceattacks, zhang2025softselectivedataobfuscation}, which significantly limits their practicality and adaptability in real-world deployments. The superior performance of RAG suggests that unmodified LLMs, which rely on explicit retrieval rather than parameter updates, are inherently less prone to data memorization and thus can help mitigate MIA risks. 

Motivated by these observations, we leverage
the complementary strengths of knowledge-injected models (high task accuracy but higher leakage) and base models (stronger privacy but weaker specialization) %, we can 
to construct a hybrid ensemble defense. We instantiate this idea as Ensemble Privacy Defense (EPD), a lightweight and model-agnostic inference-time defense method that obtains candidate answers from both a knowledge-injected and base model, and
leverages an LLM-as-a-judge to integrate their outputs, thereby offering strong post-processing protection against membership inference with no retraining overhead.

%% file: latex/Sections/4epd.tex
In this section, we introduce Ensemble Privacy Defense (EPD), a model-agnostic inference-time framework designed to mitigate membership inference risks in knowledge-intensive LLMs. Unlike existing defenses that require retraining or architectural modifications, EPD can be seamlessly applied to both finetuned and retrieval-augmented models without altering their internal parameters. 

As visualized in Figure~\ref{fig:framework}, the intuition is to harness the complementary strengths of a task-specific target model and a more general base model: while the target model delivers high-accuracy answers but is prone to privacy leakage, the base model offers stronger protection at the expense of task specialization. Given a query, EPD generates candidate responses from both models and employs an LLM-as-a-judge to evaluate and integrate them into a privacy-aware final output.

On a high level, let $q$ be a query and $a$ be the generated response. Let $\mathcal{M}_{\text{target}}$ denote a target model (either finetuned or RAG-based) and $\mathcal{M}_{\text{base}}$ denote a base model without task-specific training. EPD is a function $\mathcal{E}$ that produces the answer $a_{\text{final}} = \mathcal{E}(\mathcal{M}_{\text{target}}, \mathcal{M}_{\text{base}}, q)$ where $a_{\text{final}}$ exhibits lower membership leakage compared to direct outputs from $\mathcal{M}_{\text{target}}$.

\paragraph{Candidate Answer Generation}
For each input query $q$, we generate two candidate answers along with their corresponding loss values:
\begin{equation}
    a_{\text{target}} = \mathcal{M}_{\text{target}}(q), ~a_{\text{base}} = \mathcal{M}_{\text{base}}(q) \nonumber 
\end{equation}
\begin{equation}
    \mathcal{L}_{\text{target}}(q)
= -\frac{1}{T_{\text{tar}}}\sum_{t=1}^{T_{\text{tar}}}
\log p_{\theta_{\text{target}}}\!\left(g^{\text{tar}}_t \mid x(q),\, g^{\text{tar}}_{<t}\right)\nonumber 
\end{equation}
 \begin{equation} 
     \mathcal{L}_{\text{base}}(q)
 = -\frac{1}{T_{\text{base}}}\sum_{t=1}^{T_{\text{base}}}
 \log p_{\theta_{\text{base}}}\!\left(g^{\text{base}}_t \mid x(q),\, g^{\text{base}}_{<t}\right)\nonumber 
 \end{equation}
where $a_{\text{target}}$ represents the high-accuracy but potentially privacy-leaking response, and $a_{\text{base}}$ represents the privacy-preserving but potentially less accurate response. The loss values $\mathcal{L}_{\text{target}}$ and $\mathcal{L}_{\text{base}}$ quantify model confidence and and potential membership leakage. The loss we report is the standard causal language modeling cross-entropy (token-level negative log-likelihood).
Let $g_{\text{target}}=(g^{\text{tar}}_1,\ldots,g^{\text{tar}}_{T_{\text{tar}}})$ be the tokens generated by the target model from $q$, and
$g_{\text{base}}=(g^{\text{base}}_1,\ldots,g^{\text{base}}_{T_{\text{base}}})$ for the base model. We truncate both $g_{\text{target}} $ and $g_\text{base}$ to same length $T_{\text{tar}} = T_{\text{base}} $. We evaluate next-token likelihood on the generated region only.

\input{latex/Tables/epd_triviaQA_SFT}

\paragraph{LLM-as-a-Judge Selection}
To combine these candidates optimally, we introduce a dedicated judge model $\mathcal{M}_{\text{judge}}$ that evaluates both responses along with their loss values and generates a privacy-aware final answer. The judge model produces the final answer $a_\text{final}$ with: 
\vspace{-1.5ex}
\[
    a_{\text{final}} = \mathcal{M}_{\text{judge}}(\phi(q, a_{\text{target}}, \mathcal{L}_{\text{target}}, a_{\text{base}}, \mathcal{L}_{\text{base}}))
\]

\vspace{-1.5ex}
\noindent where $\phi$ is a prompt formatting function. $\phi$ is designed to ensure that the final output preserves the accuracy from the target model while incorporating the privacy advantage of the base model. The specific instantiation of $\phi$ is %prompted 
in Appendix~\ref{sec:prompt}.

%% file: latex/Tables/epd_triviaQA_SFT.tex
\begin{table*}[t]
\centering

\caption{Defense results of SFT-based models against seven membership inference attacks  on Trivia-QA. Each cell reports the metric value, with the relative change (\%) from the base model (no defense) shown in parentheses.}
\label{tab:triviaqa_finetune_results}
\renewcommand{\arraystretch}{1.05}
\setlength{\tabcolsep}{1.5pt}
\scriptsize   

\resizebox{\textwidth}{!}{
\begin{tabular}{l|
lll|lll|lll|lll}
\toprule
\multirow{2}{*}{Defense Method} 
& \multicolumn{3}{c|}{\textbf{SPV}} 
& \multicolumn{3}{c|}{\textbf{LiRA}} 
& \multicolumn{3}{c|}{\textbf{Recall}} 
& \multicolumn{3}{c}{\textbf{Log-Loss}} \\
& \multicolumn{1}{c}{AUC$\downarrow$} & \multicolumn{1}{c}{ASR$\downarrow$} & \multicolumn{1}{c|}{TPR@1$\downarrow$}
& \multicolumn{1}{c}{AUC$\downarrow$} & \multicolumn{1}{c}{ASR$\downarrow$} & \multicolumn{1}{c|}{TPR@1$\downarrow$}
& \multicolumn{1}{c}{AUC$\downarrow$} & \multicolumn{1}{c}{ASR$\downarrow$} & \multicolumn{1}{c|}{TPR@1$\downarrow$}
& \multicolumn{1}{c}{AUC$\downarrow$} & \multicolumn{1}{c}{ASR$\downarrow$} & \multicolumn{1}{c}{TPR@1$\downarrow$} \\
\midrule
\textit{No Defense}
& \multicolumn{1}{c}{0.550} & \multicolumn{1}{c}{0.720} & \multicolumn{1}{c|}{0.000}
& \multicolumn{1}{c}{0.452} & \multicolumn{1}{c}{0.508} & \multicolumn{1}{c|}{0.010}
& \multicolumn{1}{c}{0.753} & \multicolumn{1}{c}{0.697} & \multicolumn{1}{c|}{0.212}
& \multicolumn{1}{c}{0.763} & \multicolumn{1}{c}{0.707} & \multicolumn{1}{c}{0.135} \\
\midrule
LLM HAMP
& \underline{0.480}(-13) & \underline{0.400}(-44) & \underline{0.000}(0)
& 0.239(-49) & 0.503(-1) & 0.005(-50)
& 0.718(-4) & 0.692(-1) & 0.057(-71)
& 0.674(-12) & 0.637(-10) & 0.083(-39) \\
DP-Logits
& \underline{0.480}(-13) & 0.650(-10) & \underline{0.000}(0)
& 0.239(-49) & 0.500(-2) & 0.030(+189)
& 0.700(-7) & 0.663(-5) & 0.119(-44)
& 0.674(-12) & 0.637(-10) & 0.083(-39) \\
\textbf{EPD (Ours)}
& \underline{0.480}(-13) & 0.510(-29) & \underline{0.000}(0)
& \underline{0.230}(-49) & \underline{0.500}(-2) & \underline{0.000}(-100)
& \underline{0.613}(-19) & \underline{0.593}(-15) & \underline{0.010}(-95)
& \underline{0.601}(-21) & \underline{0.580}(-18) & \underline{0.021}(-85) \\
\bottomrule
\end{tabular}
}
%\vspace{0.8em}

\resizebox{\textwidth}{!}{
\begin{tabular}{l|
lll|lll|lll|lll}
\toprule
\multirow{2}{*}{Defense Method} 
& \multicolumn{3}{c|}{\textbf{Zlib}} 
& \multicolumn{3}{c|}{\textbf{Min-K}} 
& \multicolumn{3}{c|}{\textbf{Min-K++}} 
& \multicolumn{3}{c}{\textbf{Average (across the 7 MIAs)}} \\
& \multicolumn{1}{c}{AUC$\downarrow$} & \multicolumn{1}{c}{ASR$\downarrow$} & \multicolumn{1}{c|}{TPR@1$\downarrow$}
& \multicolumn{1}{c}{AUC$\downarrow$} & \multicolumn{1}{c}{ASR$\downarrow$} & \multicolumn{1}{c|}{TPR@1$\downarrow$}
& \multicolumn{1}{c}{AUC$\downarrow$} & \multicolumn{1}{c}{ASR$\downarrow$} & \multicolumn{1}{c|}{TPR@1$\downarrow$}
& \multicolumn{1}{c}{AUC$\downarrow$} & \multicolumn{1}{c}{ASR$\downarrow$} & \multicolumn{1}{c}{TPR@1$\downarrow$} \\
\midrule
\textit{No Defense}
& \multicolumn{1}{c}{0.719} & \multicolumn{1}{c}{0.674} & \multicolumn{1}{c|}{0.093}
& \multicolumn{1}{c}{0.692} & \multicolumn{1}{c}{0.655} & \multicolumn{1}{c|}{0.047}
& \multicolumn{1}{c}{0.780} & \multicolumn{1}{c}{0.731} & \multicolumn{1}{c|}{0.083}
& \multicolumn{1}{c}{0.673} & \multicolumn{1}{c}{0.670} & \multicolumn{1}{c}{0.083} \\
\midrule
LLM HAMP
& 0.673(-7) & 0.635(-6) & 0.078(-17)
& \underline{0.565}(-18) & \underline{0.554}(-17) & 0.031(-33)
& \underline{0.495}(-37) & \underline{0.521}(-29) & 0.021(-75)
& 0.549(-18) & 0.563(-16) & 0.039(-53) \\
DP-Logits
& 0.673(-7) & 0.640(-5) & 0.078(-17)
& \underline{0.565}(-18) & 0.557(-11) & 0.031(-33)
& 0.497(-36) & 0.533(-27) & 0.026(-69)
& 0.547(-19) & 0.597(-11) & 0.052(-37) \\
\textbf{EPD (Ours)}
& \underline{0.646}(-10) & \underline{0.614}(-9) & \underline{0.026}(-72)
& 0.589(-15) & 0.573(-13) & \underline{0.026}(-44)
& 0.519(-34) & 0.550(-25) & \underline{0.017}(-80)
& {\textbf{0.525}}(-22) & {\textbf{0.560}}(-16) & {\textbf{0.014}}(-83) \\
\bottomrule
\end{tabular}
}
\vskip -0.5ex
\end{table*}

%% file: latex/Sections/5experiments.tex
In this section, we conduct a comprehensive evaluation of the effectiveness of our proposed Ensemble Privacy Defense (EPD) framework.
\subsection{Experimental Setup}

\paragraph{Implementation} 
We follow the same methodology outlined in Section~\ref{robustness:implementation} for both SFT and RAG. The judge model $\mathcal{M}_\text{judge}$ leverages the DeepSeek-R1-8B~\cite{guo2025deepseek}. Detailed implementation specifications for all models and experimental configurations are provided in Appendix~\ref{sec:models_implementation}.

\paragraph{Datasets}
Here we evaluate 
on TriviaQA~\cite{joshi2017triviaqalargescaledistantly} and AG News~\cite{zhang2015characterlevel} datasets. Specifically, for AG News, only SFT is evaluated because AG News does not provide retrieved/relevant contexts in its original dataset. 

\paragraph{Evaluation Metrics}
Besides AUC and ASR, we also use TPR\@1\%FPR (shortened to TPR@1 for brevity) for MIA defense evaluation. 
TPR\@1\%FPR measures the true Positive Rate at 1\% False Positive Rate, indicating the attack performance under strict privacy constraints. 

\subsection{Inference-time Baselines}
We include two state-of-the-art baselines for MIA defense. Although recent research has proposed more advanced defense strategies for RAG~\cite{Anderson_2025, liu2025maskbasedmembershipinferenceattacks}, we do not include them in our experiment because the MIAs are designed for both SFT and RAG. Moreover, to ensure a fair comparison with our proposed approach, we restrict to inference-time methods that do not require retraining or architectural modifications. 

\noindent \textit{HAMP (Hiding Auxiliary Membership Privacy): } An inference-time defense method that enforces less confident predictions by introducing high-entropy soft labels and an entropy-based regularizer~\cite{chen2023overconfidence}.

\noindent \textit{DP-Logits:} Applies differential privacy to the output logits during inference by adding calibrated noise to the 
predictions~\cite{rahimian2020sampling}. 

\subsection{Main Results}
In this section, we present our experiment results against the seven representative MIA attack methods introduced in Section~\ref{sec:mia_attack_methods}. The results are presented in Tables~\ref{tab:triviaqa_finetune_results}, \ref{tab:triviaqa_rag_results}, and \ref{tab:agnews_results}.
At the same time, we present our experiment results measuring the final answer accuracy to show that our defense method preserves answer quality in Table ~\ref{tab:model_comparison}.

\input{latex/Tables/epd_triviaQA_RAG}

\subsubsection{MIA Success Rate Reduction}
As shown in Table~\ref{tab:triviaqa_finetune_results}, on TriviaQA with finetuning, our LLM Judge framework substantially reduces both AUC and ASR across most attack methods. The strongest gains are achieved on reference-based attacks: LiRA (49.2\% reduction), Log-Loss (21.2\%), Recall (18.5\%), and Zlib (10.2\%). For ASR, the method achieves notable reductions of 18.0\% on Log-Loss and 14.9\% on Recall, while maintaining competitive results on other MIAs. Averaged across metrics, EPD outperforms the runner-up by 36.2\% on robustness improvement.

Table~\ref{tab:triviaqa_rag_results} presents results on TriviaQA with RAG. Although RAG already exhibits lower susceptibility to MIAs compared to finetuned models, our framework still improves robustness further, especially on LiRA and Zlib attacks, with 100\% and 75\% TPR@1 reduction, respectively. These results highlight that for RAG settings where vulnerability is already reduced, EPD can still effectively improve the privacy robustness. Averaged across metrics, EPD outperforms the runner-up by 232.2\% on robustness improvement.

For AG News classification tasks, reported in Table~\ref{tab:agnews_results}, our framework consistently improves privacy protection across most MIAs. Notably, on the TPR@1\%FPR metric, the LLM Judge achieves large reductions, including 67\% for LiRA attacks and 51\% for Min-K++ attacks. Averaged across metrics, EPD outperforms the runner-up by 138.7\% on robustness improvement.

Overall, when compared to baseline defense methods across all three tables, our LLM Judge framework, EPD, nearly always outperforms or matches their performance across the MIA scenarios. Across the datasets, EPD enhances AUC by 205.2\%, ASR by 23.7\%, and TPR@1 by 166.8\%. These results demonstrate that the LLM Judge framework provides a robust defense against membership inference attacks, making it a practical solution for privacy-preserving LLM deployment in sensitive applications.

\subsubsection{Final Answer Accuracy}
As shown in Table~\ref{tab:model_comparison}, we evaluated the model under SFT settings on TriviaQA using Exact Match (EM) and F1-Score. We run the experiments on 1000 extracted data from TriviaQA's test set compared to the provided ground truth. We report the result in the following table and observe that EPD preserves utility, with EM/F1 fluctuations within a narrow range compared to the original target model and two other defense baselines.  

\input{latex/Tables/ablation_accuracy}

\subsection{Ablation Studies}
In this section, we conduct an extensive ablation study to investigate %the proposed EPD components. 
EPD's components. 
We will first explore the impact of Judge Model's capacity. Then we will explore the impact of $k$ in RAG retrieval on the MIA defense. Additionally, we studied the Judge model's behavior based on answers from Target model and Base model to measure if it has bias towards any of them. Finally, we study adding 
adaptive noise injection~\cite{wang2025privacy} as an additional defense mechanism for EPD.

\input{latex/Tables/epd_agnews_SFT}

\subsubsection{Impact of Judge Model Capacity.}

To investigate the dependency on model capacity, we conduct a %an ablation 
study comparing two different judge models with varying parameter scales: DeepSeek-R1-1.5B 
and DeepSeek-R1-8B. Both models are based on the DeepSeek-R1 architecture but differ %significantly 
in their parameter count and computational capacity.

The judge model's role in our framework is to evaluate candidate answers from both the target (finetuned/RAG) and base models, then synthesize a final response that balances accuracy with privacy protection. This %task 
requires sophisticated reasoning %. We 
and we hypothesize that judge models with greater parameter capacity will demonstrate superior performance in this complex decision-making process. Our experimental setup maintains identical configurations for all other %components of the framework, 
framework components, 
with the judge model capacity being the only variable.

Table~\ref{tab:judge_capacity_ablation} presents the comprehensive results comparing the two judge model variants across multiple MIA attack methods and evaluation metrics. The results demonstrate a clear correlation between judge model capacity and defense performance across all attack scenarios.

\input{latex/Tables/ablation_judge_model}

The results reveal that the larger 8B parameter model consistently outperforms the 1.5B parameter model across all attack methods and evaluation metrics except ASR in SPV-MIA method. The superior performance of the larger judge model can be attributed to several factors. Detailed analysis is provided in Appendix~\ref{sec:model_analysis}.

Our results suggest that continued scaling of judge model capacity may yield further improvements in MIA defense effectiveness, though the diminishing returns observed in other domains~\cite{hoffmann2022training, kaplan2020scaling} may also apply here. Future work should explore the optimal balance between model capacity and performance.

\subsubsection{Impact of Top-K Retrieval on RAG-based MIA Defense}
The number of retrieved contexts ($k$) directly determines how much external knowledge a RAG model conditions on, which in turn can influence its privacy behavior. To examine this relationship, we conduct a systematic ablation study analyzing how different top-$k$ retrieval settings affect MIA defense effectiveness in RAG systems.

\begin{figure}[t]
    \centering
    \begin{subfigure}[t]{0.7\linewidth}
        \centering
        \includegraphics[width=0.95\linewidth]{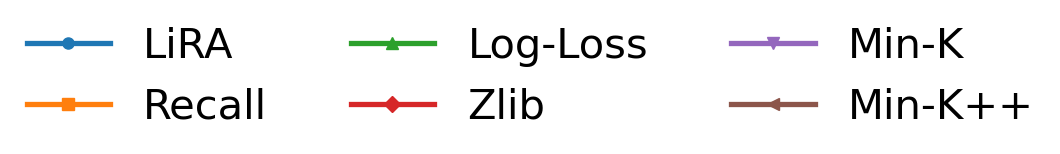}
    \end{subfigure}

    \begin{subfigure}[t]{0.8\linewidth}
        \centering
        \includegraphics[width=\linewidth]{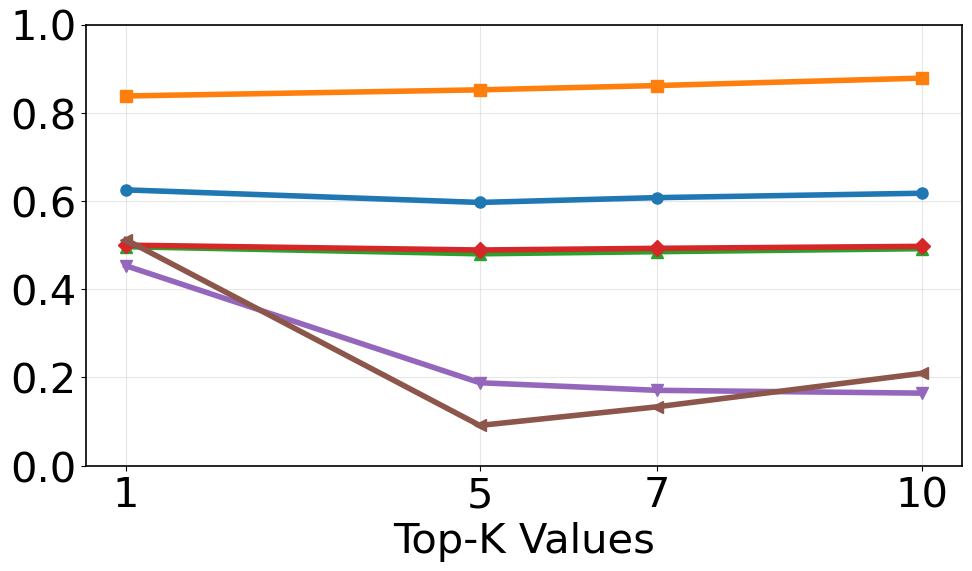}
        \caption{AUC}
        \label{fig:topkauc}
    \end{subfigure}
    \hfill

    \begin{subfigure}[t]{0.8\linewidth}
        \centering
        \includegraphics[width=\linewidth]{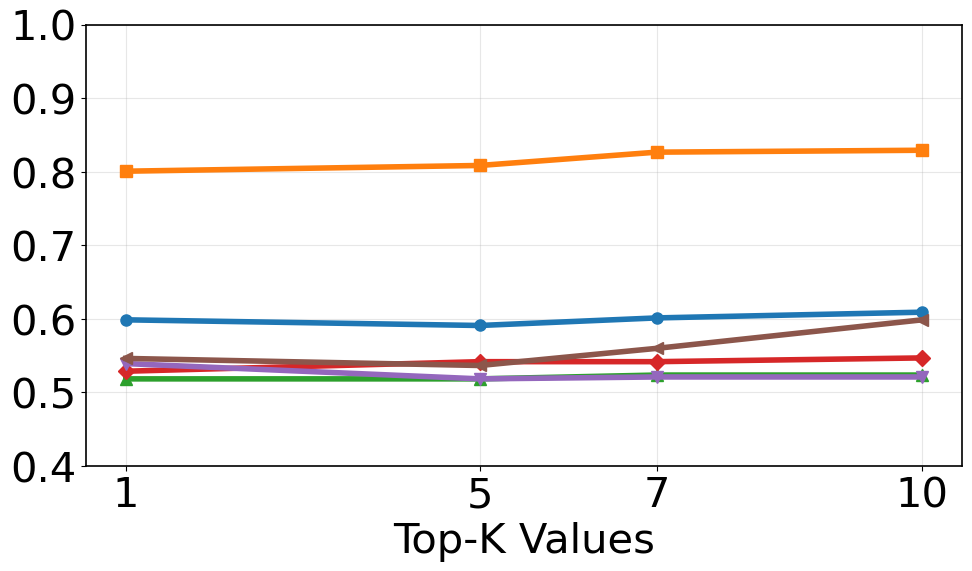}
        \caption{ASR}
        \label{fig:topkasr}
    \end{subfigure}
   
    \caption{Effect of different choices of $K$ on the attack performance. Colored lines are the attack methods.}
    \label{fig:topk_comparison}

\end{figure}

Specifically, we evaluate four retrieval configurations, $k \in {1, 5, 7, 10}$, on the SQuAD dataset using our EPD framework under the RAG setting. Figure~\ref{fig:topk_comparison} summarizes the results across multiple MIA attack methods and evaluation metrics. Overall, EPD remains robust across different values of $k$. While we observe a slight increase in ASR as $k$ grows, which aligns with intuition, since retrieving more passages can help the target model generate more grounded responses, the magnitude of this increase is small, highlighting the stability of our defense against most realistic retrieval depth.

\subsubsection{Judge Model Bias}
%In our design of EPD, 
In EPD it is assumed that the judge model is not designed to reward stylistic patterns, verbosity, or structural preferences from either candidate model. Instead, its role is to aggregate outputs while suppressing signals indicative of membership data. To directly assess whether the judge systematically favors either the target or the base model, we conducted an explicit post-hoc bias evaluation. 

As shown in detail within Algorithm~\ref{alg:judge_categorization} of Appendix ~\ref{appendix:pseudo}, each judge decision was categorized into Target, Base, and Mixed through a hierarchical matching process. To show that the Judge model does not bias towards certain model outputs, we compute the percentage of answers that the judge model directly chooses the target model output (19.9\%), base model output (10.4\%), and the percentage of mixed output (69.7\%). Additionally, we compute the bias score as the absolute difference between the percentage of instances in which the judge selected the target answer versus the base answer (9.5\%). As evident by the results, the judge predominantly produces mixed or hybrid responses, which aligns with the intended design goal of blending information across heterogeneous distributions rather than copying either model’s style. This empirical distribution further confirms that the judge behaves as a neutral aggregator.

\subsubsection{Impact of Adaptive Noise Injection} 

Building upon recent advances in privacy-aware decoding for MIA defense~\cite{wang2025privacy}, we evaluate adaptive noise injection as an additional defense mechanism in our LLM Judge framework. The key intuition is that strategic perturbations introduced during generation may further obfuscate membership signals without significantly degrading answer quality. We adapt the Privacy-Aware Decoding (PAD) methodology to our ensemble setting by modulating noise strength using similarity and loss differences between the target and base model responses, with full details provided in Appendix~\ref{sec:noise_additon}. Table~\ref{tab:noise_injection_ablation_transposed} reports the performance impact across six MIA attacks under two noise thresholds. The results show that noise injection has highly attack-dependent behavior: LiRA benefits consistently across both noise levels, Recall and Log-Loss exhibit minor improvements, Zlib shows minimal change, and Min-K/Min-K++ even degrade under certain thresholds. These heterogeneous outcomes highlight that noise-based defenses interact differently with score distributions across attacks and are therefore not universally effective.

Overall, while the more aggressive threshold ($\tau=0.05$) sometimes enhances privacy metrics, its impact remains inconsistent and occasionally counterproductive. These findings indicate that adaptive noise injection offers only modest additional protection on top of the ensemble-based judge design, which already suppresses membership-specific signals effectively. In practice, this suggests that the LLM Judge’s inherent aggregation behavior serves as the primary driver of privacy improvements, with noise playing only a secondary and context-dependent role. As such, noise perturbation may be best viewed as an optional enhancement rather than a core component of the defense pipeline.

\begin{table}[t]
\large
\centering
\caption{Impact of Adaptive Noise Injection on LLM Judge MIA Defense Performance (Transposed, Two-Decimal Precision)}
\label{tab:noise_injection_ablation_transposed}
\setlength{\tabcolsep}{1.5pt}

\resizebox{\columnwidth}{!}{%
\begin{tabular}{l|ccc|ccc|ccc}
\toprule
\multirow{2}{*}{Methods} 
& \multicolumn{3}{c|}{LLM Judge (Baseline)} 
& \multicolumn{3}{c|}{+ Noise ($\tau=0.1$)} 
& \multicolumn{3}{c}{+ Noise ($\tau=0.05$)} \\
\cmidrule{2-10}
& AUC & ASR & TPR@1 & AUC & ASR & TPR@1 & AUC & ASR & TPR@1 \\
\midrule
LiRA      & 0.23 & 0.50 & 0.00 & 0.22 & 0.50 & 0.01 & 0.22 & 0.50 & 0.00 \\
Recall    & 0.61 & 0.60 & 0.03 & 0.63 & 0.61 & 0.03 & 0.62 & 0.60 & 0.02 \\
Log-Loss  & 0.61 & 0.61 & 0.03 & 0.63 & 0.61 & 0.03 & 0.62 & 0.61 & 0.02 \\
Zlib      & 0.65 & 0.61 & 0.03 & 0.66 & 0.66 & 0.02 & 0.65 & 0.61 & 0.03 \\
Min-K     & 0.60 & 0.59 & 0.03 & 0.65 & 0.65 & 0.01 & 0.63 & 0.61 & 0.05 \\
Min-K++   & 0.52 & 0.55 & 0.01 & 0.54 & 0.54 & 0.06 & 0.55 & 0.59 & 0.02 \\
\bottomrule
\end{tabular}%
}
\end{table}

%% file: latex/Tables/epd_triviaQA_RAG.tex
\begin{table*}[t]
\centering
%\vspace{-0.5ex}
\caption{Defense results of RAG-based models against seven membership inference attacks  on Trivia-QA.}
\label{tab:triviaqa_rag_results}
\renewcommand{\arraystretch}{1.05}
\setlength{\tabcolsep}{1.5pt}
\footnotesize  

\resizebox{\textwidth}{!}{
\begin{tabular}{l|
lll|lll|lll|lll}
\toprule
\multirow{2}{*}{Methods} 
& \multicolumn{3}{c|}{\textbf{SPV}} 
& \multicolumn{3}{c|}{\textbf{LiRA}} 
& \multicolumn{3}{c|}{\textbf{Recall}} 
& \multicolumn{3}{c}{\textbf{Log-Loss}} \\
& \multicolumn{1}{c}{AUC$\downarrow$} & \multicolumn{1}{c}{ASR$\downarrow$} & \multicolumn{1}{c|}{TPR@1$\downarrow$}
& \multicolumn{1}{c}{AUC$\downarrow$} & \multicolumn{1}{c}{ASR$\downarrow$} & \multicolumn{1}{c|}{TPR@1$\downarrow$}
& \multicolumn{1}{c}{AUC$\downarrow$} & \multicolumn{1}{c}{ASR$\downarrow$} & \multicolumn{1}{c|}{TPR@1$\downarrow$}
& \multicolumn{1}{c}{AUC$\downarrow$} & \multicolumn{1}{c}{ASR$\downarrow$} & \multicolumn{1}{c}{TPR@1$\downarrow$} \\
\midrule
\textit{No Defense}
& \multicolumn{1}{c}{0.491} & \multicolumn{1}{c}{0.580} & \multicolumn{1}{c|}{0.005}
& \multicolumn{1}{c}{0.377} & \multicolumn{1}{c}{0.513} & \multicolumn{1}{c|}{0.010}
& \multicolumn{1}{c}{0.386} & \multicolumn{1}{c}{0.603} & \multicolumn{1}{c|}{0.031}
& \multicolumn{1}{c}{0.599} & \multicolumn{1}{c}{0.662} & \multicolumn{1}{c}{0.104} \\
\midrule
LLM HAMP
& \underline{0.480}(-2) & \underline{0.400}(-31) & \underline{0.000}(-100)
& 0.270(-28) & 0.500(-2) & 0.010(0)
& 0.652(+69) & 0.619(+3) & 0.036(+17)
& 0.669(+12) & 0.635(-4) & 0.062(-40) \\
DP-Logits
& \underline{0.480}(-2) & 0.650(+12) & \underline{0.000}(-100)
& 0.270(-28) & 0.500(-2) & 0.030(+189)
& 0.680(+76) & 0.637(+6) & 0.042(+33)
& 0.669(+12) & 0.635(-4) & 0.062(-40) \\
\textbf{EPD (Ours)}
& \underline{0.480}(-2) & 0.510(-12) & \underline{0.000}(-100)
& \underline{0.240}(-36) & \underline{0.500}(-2) & \underline{0.000}(-100)
& \underline{0.558}(+45) & \underline{0.557}(-8) & 0.036(+17)
& \underline{0.539}(-10) & \underline{0.552}(-17) & \underline{0.026}(-75) \\
\bottomrule
\end{tabular}}

\resizebox{\textwidth}{!}{
\begin{tabular}{l|
lll|lll|lll|lll}
\toprule
\multirow{2}{*}{Methods} 
& \multicolumn{3}{c|}{\textbf{Zlib}} 
& \multicolumn{3}{c|}{\textbf{Min-K}} 
& \multicolumn{3}{c|}{\textbf{Min-K++}} 
& \multicolumn{3}{c}{\textbf{Average (across the 7 MIAs)}} \\
& \multicolumn{1}{c}{AUC$\downarrow$} & \multicolumn{1}{c}{ASR$\downarrow$} & \multicolumn{1}{c|}{TPR@1$\downarrow$}
& \multicolumn{1}{c}{AUC$\downarrow$} & \multicolumn{1}{c}{ASR$\downarrow$} & \multicolumn{1}{c|}{TPR@1$\downarrow$}
& \multicolumn{1}{c}{AUC$\downarrow$} & \multicolumn{1}{c}{ASR$\downarrow$} & \multicolumn{1}{c|}{TPR@1$\downarrow$}
& \multicolumn{1}{c}{AUC$\downarrow$} & \multicolumn{1}{c}{ASR$\downarrow$} & \multicolumn{1}{c}{TPR@1$\downarrow$} \\
\midrule
\textit{No Defense}
& \multicolumn{1}{c}{0.686} & \multicolumn{1}{c}{0.619} & \multicolumn{1}{c|}{0.104}
& \multicolumn{1}{c}{0.690} & \multicolumn{1}{c}{0.595} & \multicolumn{1}{c|}{0.036}
& \multicolumn{1}{c}{0.780} & \multicolumn{1}{c}{0.557} & \multicolumn{1}{c|}{0.041}
& \multicolumn{1}{c}{0.571} & \multicolumn{1}{c}{0.590} & \multicolumn{1}{c}{0.047} \\
\midrule
LLM HAMP
& 0.669(-3) & 0.635(+3) & 0.104(0)
& 0.598(-13) & 0.588(-1) & 0.031(-14)
& 0.580(-26) & 0.575(+3) & 0.005(-88)
& 0.560(-2) & 0.565(-4) & 0.035(-25) \\
DP-Logits
& 0.667(-3) & 0.635(+3) & 0.104(0)
& 0.598(-13) & 0.586(-2) & 0.021(-43)
& 0.577(-26) & 0.567(+2) & 0.005(-88)
& 0.563(-2) & 0.601(+2) & 0.038(-21) \\
\textbf{EPD (Ours)}
& \underline{0.616}(-10) & \underline{0.596}(-4) & \underline{0.067}(-35)
& \underline{0.585}(-15) & \underline{0.580}(-3) & \underline{0.016}(-57)
& \underline{0.505}(-35) & \underline{0.541}(-3) & \underline{0.010}(-75)
& \textbf{0.503}(-12) & \textbf{0.548}(-7) & \textbf{0.022}(-53) \\
\bottomrule
\end{tabular}}
\vskip -0.5ex
\end{table*}

%% file: latex/Tables/ablation_accuracy.tex
\begin{table}[t]
\small
\centering
\caption{Accuracy Comparison of Different Models}
\label{tab:model_comparison}
\begin{tabular}{lcc} 
\toprule
\textbf{Model} & \textbf{Exact Match} & \textbf{F1-Score} \\
\midrule 
Ground Truth    & 0.3560 & 0.4466 \\ \midrule
DP-Logits       & 0.2690 & 0.3150 \\
LLM-Hamp        & 0.2070 & 0.2874 \\
EPD-Judge Model & \textbf{0.3320} & \textbf{0.4047} \\
\bottomrule 
\end{tabular}
\end{table}

%% file: latex/Tables/epd_agnews_SFT.tex
\begin{table*}[t]
\centering
\caption{Defense results of SFT-based models against seven membership inference attacks on Ag News.}
\label{tab:agnews_results}
\renewcommand{\arraystretch}{1.05}
\setlength{\tabcolsep}{1.5pt}
\tiny 
\vspace{-0.5ex}

\resizebox{\textwidth}{!}{
\begin{tabular}{l|
lll|lll|lll|lll}
\toprule
\multirow{2}{*}{Methods} 
& \multicolumn{3}{c|}{\textbf{SPV}} 
& \multicolumn{3}{c|}{\textbf{LiRA}} 
& \multicolumn{3}{c|}{\textbf{Recall}} 
& \multicolumn{3}{c}{\textbf{Log-Loss}} \\
& \multicolumn{1}{c}{AUC$\downarrow$} & \multicolumn{1}{c}{ASR$\downarrow$} & \multicolumn{1}{c|}{TPR@1$\downarrow$}
& \multicolumn{1}{c}{AUC$\downarrow$} & \multicolumn{1}{c}{ASR$\downarrow$} & \multicolumn{1}{c|}{TPR@1$\downarrow$}
& \multicolumn{1}{c}{AUC$\downarrow$} & \multicolumn{1}{c}{ASR$\downarrow$} & \multicolumn{1}{c|}{TPR@1$\downarrow$}
& \multicolumn{1}{c}{AUC$\downarrow$} & \multicolumn{1}{c}{ASR$\downarrow$} & \multicolumn{1}{c}{TPR@1$\downarrow$} \\
\midrule
\textit{No Defense}
& \multicolumn{1}{c}{0.490} & \multicolumn{1}{c}{0.580} & \multicolumn{1}{c|}{0.005}
& \multicolumn{1}{c}{0.529} & \multicolumn{1}{c}{0.513} & \multicolumn{1}{c|}{0.027}
& \multicolumn{1}{c}{0.513} & \multicolumn{1}{c}{0.603} & \multicolumn{1}{c|}{0.012}
& \multicolumn{1}{c}{0.514} & \multicolumn{1}{c}{0.662} & \multicolumn{1}{c}{0.013} \\
\midrule
LLM HAMP
& \underline{0.480}(-2) & \underline{0.400}(-31) & \underline{0.000}(-100)
& 0.525(-1) & 0.530(+3) & 0.026(-4)
& 0.509(-1) & \underline{0.508}(-16) & \underline{0.011}(-7)
& 0.514(0) & 0.510(-23) & \underline{0.012}(-6) \\
DP-Logits
& \underline{0.480}(-2) & 0.650(+12) & \underline{0.000}(-100)
& 0.525(-1) & 0.530(+3) & 0.026(-4)
& 0.514(0) & 0.511(-15) & 0.012(-1)
& 0.514(0) & 0.510(-23) & \underline{0.012}(-6) \\
\textbf{EPD (Ours)}
& \underline{0.480}(-2) & 0.510(-12) & \underline{0.000}(-100)
& \underline{0.495}(-7) & \underline{0.501}(-2) & \underline{0.009}(-68)
& \underline{0.502}(-2) & 0.510(-15) & 0.012(-3)
& \underline{0.501}(-3) & \underline{0.510}(-23) & 0.013(-2) \\
\bottomrule
\end{tabular}}

\resizebox{\textwidth}{!}{
\begin{tabular}{l|
lll|lll|lll|lll}
\toprule
\multirow{2}{*}{Methods} 
& \multicolumn{3}{c|}{\textbf{Zlib}} 
& \multicolumn{3}{c|}{\textbf{Min-K}} 
& \multicolumn{3}{c|}{\textbf{Min-K++}} 
& \multicolumn{3}{c}{\textbf{Average (across the 7 MIAs)}} \\
& \multicolumn{1}{c}{AUC$\downarrow$} & \multicolumn{1}{c}{ASR$\downarrow$} & \multicolumn{1}{c|}{TPR@1$\downarrow$}
& \multicolumn{1}{c}{AUC$\downarrow$} & \multicolumn{1}{c}{ASR$\downarrow$} & \multicolumn{1}{c|}{TPR@1$\downarrow$}
& \multicolumn{1}{c}{AUC$\downarrow$} & \multicolumn{1}{c}{ASR$\downarrow$} & \multicolumn{1}{c|}{TPR@1$\downarrow$}
& \multicolumn{1}{c}{AUC$\downarrow$} & \multicolumn{1}{c}{ASR$\downarrow$} & \multicolumn{1}{c}{TPR@1$\downarrow$} \\
\midrule
\textit{No Defense}
& \multicolumn{1}{c}{0.515} & \multicolumn{1}{c}{0.619} & \multicolumn{1}{c|}{0.012}
& \multicolumn{1}{c}{0.514} & \multicolumn{1}{c}{0.595} & \multicolumn{1}{c|}{0.012}
& \multicolumn{1}{c}{0.514} & \multicolumn{1}{c}{0.557} & \multicolumn{1}{c|}{0.017}
& \multicolumn{1}{c}{0.513} & \multicolumn{1}{c}{0.590} & \multicolumn{1}{c}{0.014} \\
\midrule
LLM HAMP
& 0.515(0) & 0.517(-17) & 0.013(+6)
& 0.513(0) & 0.514(-14) & 0.013(+7)
& 0.516(0) & 0.515(-8) & 0.015(-14)
& 0.510(-1) & \textbf{0.499}(-15) & 0.013(-9) \\
DP-Logits
& 0.515(0) & 0.516(-17) & 0.013(+6)
& 0.513(0) & 0.514(-14) & 0.012(+6)
& 0.515(0) & 0.515(-8) & 0.015(-10)
& 0.511(0) & 0.535(-9) & 0.013(-7) \\
\textbf{EPD (Ours)}
& \underline{0.507}(-2) & \underline{0.511}(-17) & \underline{0.010}(-21)
& \underline{0.495}(-4) & \underline{0.505}(-15) & \underline{0.010}(-17)
& \underline{0.500}(-3) & \underline{0.504}(-9) & \underline{0.008}(-51)
& \textbf{0.497}(-3) & 0.507(-14) & \textbf{0.009}(-38) \\
\bottomrule
\end{tabular}}
\vskip -1.5ex
\end{table*}

%% file: latex/Tables/ablation_judge_model.tex
\begin{table}[t]
\centering
\footnotesize 
\caption{Results evaluating the impact of EPD's judge model capacity towards defending MIAs. } 
\label{tab:judge_capacity_ablation}
\setlength{\tabcolsep}{2.5pt}

\begin{tabular}{l|ccc|ccc}
\toprule
\multirow{2}{*}{\begin{tabular}{c} MIA \\ Method \end{tabular}}

& \multicolumn{3}{c|}{DeepSeek-R1-1.5B} 
& \multicolumn{3}{c}{DeepSeek-R1-8B} \\
\cmidrule{2-7}
& AUC & ASR & TPR@1 & AUC & ASR & TPR@1 \\
\midrule
SPV-MIA   & 0.52   & \underline{0.32}   & 0.03   & \underline{0.48}   & 0.51   & \underline{0.00}   \\
LiRA      & 0.27 & 0.59 & 0.01 & \underline{0.23} & \underline{0.50} & \underline{0.00} \\
Recall    & 0.63 & 0.61 & 0.05 & \underline{0.61} & \underline{0.59} & \underline{0.01} \\
Log-Loss  & 0.62 & 0.61 & 0.03 & \underline{0.60} & \underline{0.58} & \underline{0.02} \\
Zlib      & 0.65 & 0.63 & 0.04 &\underline{ 0.65} & \underline{0.61} & \underline{0.03} \\
Min-K     & 0.62 & 0.59 & \underline{0.03} & \underline{0.59} & \underline{0.57} & \underline{0.03} \\
Min-K++   & 0.67 & 0.63 & \underline{0.02} & \underline{0.52} & \underline{0.55} & \underline{0.02} \\
\midrule
Average & 0.569 & 0.569 & 0.030 & \textbf{0.526} & \textbf{0.559} & \textbf{0.0157} \\
\bottomrule
\end{tabular}%
\vskip -1ex
\end{table}

%% file: latex/Sections/relatedwork.tex
Defending against Membership Inference Attacks (MIAs) in Large Language Models (LLMs) has emerged as a critical research frontier. While traditional defenses such as Differential Privacy (DP)~\cite{abadi2016deep} offer provable privacy guarantees, applying them directly to LLMs often incurs a severe "utility-privacy trade-off," resulting in significant degradation of text generation quality and reasoning capabilities. Consequently, recent literature has bifurcated into training-time and inference-time strategies.

\paragraph{Training-time Approaches.}
Training-time approaches fundamentally alter the model's learning process. Techniques ranging from strict DP-SGD to data sanitization and deduplication aim to minimize memorization at the source. Recent specific methods, such as those by Tran et al.~\cite{tran2025tokens, tran2024dual}, modify the training objective or dataset composition to enhance resistance against extraction attacks. However, these methods suffer from substantial practical limitations: they require access to the full training corpus and necessitate computationally prohibitive retraining. For modern LLMs with billions of parameters, retraining for privacy is often infeasible, rendering these defenses impractical for pre-trained, proprietary, or API-served models.

\paragraph{Inference-time Approaches}
Inference-time defenses, in contrast, operate during the generation phase, making them model-agnostic and deployment-friendly. These methods generally aim to mask the confidence signals that attackers exploit. For instance, DP-Logits~\cite{rahimian2020sampling} introduces noise to the output logits, effectively applying DP at the decoding stage to obscure the true probability distribution. Similarly, HAMP~\cite{chen2023overconfidence} mitigates membership leakage by enforcing low-confidence predictions via high-entropy soft labels and entropy-based regularization. Despite their deployment advantages, inference-time defenses remain underexplored. Current methods often struggle to balance the masking of membership signals with the preservation of semantic coherence, motivating our investigation into lightweight, effective inference-time mechanisms. (See Appendix~\ref{appendix:related} for an extended survey).

%% file: latex/Sections/6conclusion.tex
In this work, we conducted a comprehensive investigation into the privacy vulnerabilities of RAG and finetuned LLMs against membership inference attacks. Our systematic evaluation reveals that RAG models demonstrate significantly stronger resistance to MIAs compared to their finetuned counterparts, though this enhanced privacy robustness comes at the cost of increased inference latency. To address the privacy vulnerabilities in both paradigms, we introduced the \textit{Ensemble Privacy Defense (EPD)} framework, which integrates the outputs of a finetuned LLM, a non-finetuned LLM, and a dedicated judge model. Our comprehensive experiments demonstrate that this ensemble-based approach substantially reduces MIA success rates across diverse datasets and task types. 

%% file: latex/Sections/appendix.tex
\section{RAG vs Finetune}
\input{latex/Tables/sft_vs_rag}
Detailed results found in Table~\ref{tab:baseline_triviaqa_squad_comparison}. 

\section{Prompt for LLM Judge}
\label{sec:prompt}
\paragraph{Judge prompt}
\begin{quote}\small\ttfamily
You are a judge that needs to generate a final answer based on two candidate answers.\\
Consider both PRIVACY (low-loss answers that might leak training data) and ACCURACY (being factually correct).\\
\\
Question: \(q\)\\
Answer A (target): \(a_{\text{target}}\)\\
Answer B (base): \(a_{\text{base}}\)\\
\\
Model confidence (lower loss indicates higher confidence):\\
\quad -- Target answer loss: \(\mathcal{L}_{\text{target}}\)\\
\quad -- Base\ \ \ answer loss: \(\mathcal{L}_{\text{base}}\)\\
\\
Your task:\\
1. Weigh privacy and factual accuracy.\\
2. Generate a new final answer that:\\
\quad - Maintains accuracy while protecting privacy\\
\quad - Uses general language instead of specific details\\
\quad - Avoids exact quotes or precise measurements\\
\quad - Combines the best aspects of both answers\\
\\
Generate your final answer:
\end{quote}

\section{Dataset Details}
\label{sec:dataset_details}

\input{latex/Tables/dataset_statistics}

We evaluate our proposed framework on three diverse datasets to ensure comprehensive assessment across different domains and task types:

\begin{itemize}
    \item \textbf{TriviaQA (Unfiltered)}: A large-scale reading comprehension dataset containing 95K question-answer pairs derived from Wikipedia articles. The dataset features complex factual questions that require reasoning over multiple documents, making it suitable for evaluating both knowledge retrieval and generation capabilities.
    \item \textbf{SQuAD}: The Stanford Question Answering Dataset, comprising 100K+ question-answer pairs based on Wikipedia articles. SQuAD focuses on extractive question answering, where answers are spans of text from the given context, providing a different challenge compared to generative tasks.
    \item \textbf{AG News}: A news classification dataset containing 120K news articles from four categories (World, Sports, Business, Sci/Tech). This dataset represents a different task type (classification) and domain (news), allowing us to assess the generalizability of our privacy defense mechanisms.
\end{itemize}

\section{Models and Implementation Details}
\label{sec:models_implementation}

\subsection{Supervised Finetuning Setup}
We employ Llama-2-7B as the base model for supervised finetuning. To efficiently adapt the large model to our target datasets, we utilize Parameter-Efficient Fine-Tuning (PEFT) techniques, specifically LoRA (Low-Rank Adaptation), which significantly reduces the number of trainable parameters while maintaining performance. The training process incorporates early stopping to prevent overfitting and ensure optimal generalization. Key hyperparameters include a learning rate of $1\times10^{-4}$, LoRA rank of 8, LoRA alpha of 16, batch size of 4, and a block size of 128. Early stopping is applied based on validation loss to avoid overfitting.

\subsection{Retrieval-Augmented Generation Setup}
For the retrieval-augmented generation (RAG) setting, we use the original (non-finetuned) Llama-2-7B as the core LLM. For dense retrieval, we employ the SentenceTransformer~\cite{reimers2019sentencebertsentenceembeddingsusing} "all-MiniLM-L6-v2" model as the sentence encoder to embed both the input queries and all context passages from the TriviaQA and SQuAD datasets. All context embeddings are precomputed and stored. At inference time, we compute the cosine similarity between the query embedding and all stored context embeddings, retrieving the top-$k$ most similar contexts (with $k=5$) to construct the final prompt for the LLM. This setup enables efficient and effective retrieval-augmented generation for open-domain question answering.

\subsection{LLM Judge Configuration}
For our ensemble defense framework, we configure the components as follows:
\begin{itemize}[leftmargin=*]
    \item \textbf{Target Model}: The target model is the finetuned Llama-2-7B model (as described above) or original Llama-2-7B model embedded RAG model. 
    \item \textbf{Base Model}: The base model is an original Llama-2-7B model without any finetuning. 
    \item \textbf{Judge Model}: The judge model is DeepSeek-R1-Distill-Qwen-8B (i.e., a distilled model providing efficient and reliable judgment capabilities). 
\end{itemize}
\section{Model Capacity Analysis}
\label{sec:model_analysis}

The superior performance of the larger judge model can be attributed to several factors. First, increased parameter capacity enables more sophisticated understanding of complex prompts that require reasoning about privacy implications~\cite{wei2022chain, kojima2022large}. The 8B model demonstrates enhanced ability to identify potentially privacy-leaking information in candidate answers and generate appropriate alternatives that maintain utility while preserving privacy.

Second, larger models exhibit improved instruction-following capabilities, which is crucial for the judge model's task of synthesizing answers from multiple sources while adhering to privacy constraints~\cite{ouyang2022training, wei2021finetuned}. The enhanced capacity allows for more nuanced evaluation of the trade-offs between accuracy and privacy, leading to better-informed decisions about answer selection and modification.

Third, the increased model capacity facilitates better handling of edge cases and complex scenarios where simple heuristics may fail. This is particularly important for MIA defense, as attackers often exploit subtle patterns in model outputs that require sophisticated reasoning to detect and mitigate.

These findings have important implications for the practical deployment of our LLM Judge framework. While the 1.5B model provides a computationally efficient baseline, the 8B model offers significantly enhanced privacy protection at the cost of increased computational requirements. The choice between these models should be guided by the specific privacy requirements and computational constraints of the target application.

\section{Adaptive Noise Injection Strategy}
\label{sec:noise_additon}

\paragraph{Answer Similarity} We measure the semantic similarity between target and base model responses using cosine similarity of their embeddings:
    \begin{equation}
        \text{sim}(a_{\text{target}}, a_{\text{base}}) = \frac{\mathbf{e}_{a_{\text{target}}} \cdot \mathbf{e}_{a_{\text{base}}}}{||\mathbf{e}_{a_{\text{target}}}|| \cdot ||\mathbf{e}_{a_{\text{base}}}||}\nonumber 
    \end{equation}
    where $\mathbf{e}_{a_{\text{target}}}$ and $\mathbf{e}_{a_{\text{base}}}$ are the embeddings of the target and base model answers, respectively.

\paragraph{Loss Difference} We compute the normalized difference between the target and base model losses:
    \begin{equation}
        \Delta_{\text{loss}} = \frac{|\mathcal{L}_{\text{target}} - \mathcal{L}_{\text{base}}|}{\max(\mathcal{L}_{\text{target}}, \mathcal{L}_{\text{base}})}\nonumber 
    \end{equation}
    where $\mathcal{L}_{\text{target}}$ and $\mathcal{L}_{\text{base}}$ are the cross-entropy losses of the target and base models, respectively.

The noise injection strength is determined by a weighted combination of these metrics:
\begin{equation}
    \beta = w_{\text{sim}} \cdot (1 - \text{sim}(a_{\text{target}}, a_{\text{base}})) + w_{\text{loss}} \cdot \Delta_{\text{loss}}\nonumber 
\end{equation}
where $w_{\text{sim}}$ and $w_{\text{loss}}$ are weighting parameters (set to 0.6 and 0.4, respectively, in our experiments).

\paragraph{Threshold-based Noise Activation:} Noise injection is activated when the computed strength exceeds predefined thresholds. We evaluate two threshold configurations: $\tau_1 = 0.1$ and $\tau_2 = 0.05$, representing conservative and aggressive noise injection strategies, respectively. When $\beta > \tau$, the system applies noise injection to the final answer generation process.

\paragraph{Token-level Noise Implementation:} Following the PAD methodology~\cite{wang2025privacy}, we implement token-level noise injection for tokens that appear in both the target answer and the final judge-generated answer. For each token $t$ at position $i$, we inject calibrated Gaussian noise into the logits:
\begin{equation}
    \tilde{\mathbf{z}}_i = \mathbf{z}_i + \mathcal{N}(0, \sigma_i^2 \mathbf{I})\nonumber 
\end{equation}
where $\mathbf{z}_i$ represents the original logits for token $i$, and $\sigma_i$ is the adaptive noise scale determined by:
\begin{equation}
    \sigma_i = \sigma_{\text{base}} \cdot \lambda_{\text{amp}} \cdot \beta \cdot \exp(-\alpha \cdot \text{confidence}_i)\nonumber 
\end{equation}
where $\sigma_{\text{base}}$ is the base noise scale, $\lambda_{\text{amp}}$ is the amplification factor, $\alpha$ is the confidence decay parameter, and $\text{confidence}_i$ is the model's confidence for token $i$ measured by the logit margin between the top two predictions.

To provide formal privacy guarantees, we employ a Rényi Differential Privacy (RDP) accountant that tracks the cumulative privacy loss across all noise-injected tokens. For each noise injection step, the RDP cost is computed as:
\vspace{-1ex}
\begin{equation}
    \text{RDP}_{\alpha}(\mathcal{M}) = \frac{\alpha \sigma_i^2}{2}\nonumber 
\end{equation}

\vspace{-1ex}
\noindent where $\alpha > 1$ is the RDP order parameter. The total privacy cost is accumulated across all protected tokens, providing an explicit $(\varepsilon, \delta)$-differential privacy guarantee for the ensemble response.

\section{Extended Related Work}
\label{appendix:related}
\paragraph{MIA Defense Mechanism}
Defending against MIAs in LLMs is an active area of research. Traditional defenses such as differential privacy (DP)~\cite{abadi2016deep} provide provable privacy guarantees but often at the cost of significant utility degradation. Practical defenses aim to empirically reduce membership leakage while maintaining model performance.

Recent work on LLM MIA defense can be broadly categorized into training-time and inference-time approaches. Training-time methods, such as those proposed by Tran et al.~\cite{tran2025tokens, tran2024dual}, modify the training process and dataset to enhance privacy protection. However, these approaches require retraining pre-trained models, which becomes computationally prohibitive and impractical for large-scale LLMs given their massive parameter counts and training costs. In contrast, inference-time defenses operate without modifying model parameters or requiring retraining, making them more suitable for real-world deployment. HAMP~\cite{chen2023overconfidence} enforces less confident predictions by introducing high-entropy soft labels and an entropy-based regularizer, while DP-Logits applies differential privacy to the output logits during inference. Rahimian et al.~\cite{rahimian2020sampling} propose sampling-based defenses that modify outputs at inference time to reduce membership leakage. These inference-time approaches provide a more practical solution for privacy protection in large-scale LLM deployments.

Despite the advantages of inference-time defenses, they remain relatively underexplored compared to training-time approaches. This gap motivates our investigation into more effective inference-time defense mechanisms that can provide robust privacy protection without the computational overhead of model retraining.

\section{Algorithm Pseudo Code}
\label{appendix:pseudo}
\input{latex/Tables/bias_algo}

%% file: latex/Tables/sft_vs_rag.tex
\begin{table*}[t]
\centering
\caption{Comparison of Robustness Evaluation Results on Trivia-QA and SQuAD (RAG vs. SFT)}
\label{tab:baseline_triviaqa_squad_comparison}
\renewcommand{\arraystretch}{1.05}
\resizebox{\textwidth}{!}{%
\begin{tabular}{l|l|cc|cc|cc|cc|cc|cc|cc}
\toprule
\multirow{2}{*}{Dataset} & \multirow{2}{*}{Paradigm}
  & \multicolumn{2}{c|}{SPV-MIA}
  & \multicolumn{2}{c|}{LiRA}
  & \multicolumn{2}{c|}{Recall}
  & \multicolumn{2}{c|}{Log-Loss}
  & \multicolumn{2}{c|}{Zlib}
  & \multicolumn{2}{c|}{Min-K}
  & \multicolumn{2}{c}{Min-K++} \\
\cline{3-16}
& & AUC & ASR & AUC & ASR & AUC & ASR & AUC & ASR & AUC & ASR & AUC & ASR & AUC & ASR \\
\midrule
\multirow{2}{*}{Trivia-QA} 
  & SFT 
  & 0.5454  & 0.7210   
  & 0.4521 & \textbf{0.5078}
  & 0.7526 & 0.6969 
  & 0.7628 & 0.7073 
  & 0.7187 & 0.6736 
  & 0.6918 & 0.6554 
  & 0.7804 & 0.7306 \\
  & RAG 
  & \textbf{0.4906} & \textbf{0.5090}  
  & \textbf{0.3768} & 0.5126 
  & \textbf{0.3856} & \textbf{0.6026} 
  & \textbf{0.5991} & \textbf{0.6618} 
  & \textbf{0.6863} & \textbf{0.6192}
  & \textbf{0.6900} & \textbf{0.5951} 
  & \textbf{0.7800} & \textbf{0.5570} \\
\midrule
\multirow{2}{*}{SQuAD} 
  & SFT 
  & \textbf{0.4951} & \textbf{0.3100} 
  & 0.8380 & 0.8005 
  & 0.4964 & 0.5181 
  & 0.6252 & 0.5984 
  & 0.4999 & 0.5285 
  & \textbf{0.4533} & 0.5389 
  & \textbf{0.5123} & \textbf{0.5259} \\
  & RAG 
  & 0.5025 & 0.3500 
  & \textbf{0.6149} & \textbf{0.6088} 
  & \textbf{0.4102} & \textbf{0.5130}
  & \textbf{0.4546} & \textbf{0.5233} 
  & \textbf{0.4523} & \textbf{0.5233} 
  & 0.4779 & \textbf{0.5285} 
  & 0.5145 & 0.5440 \\
\midrule
\multirow{2}{*}{Average} 
  & SFT 
  & 0.5203 & 0.5155 
  & 0.6450 & 0.6542 
  & 0.6245 & 0.6075 
  & 0.6940 & 0.6529 
  & 0.6093 & 0.6011 
  & \textbf{0.5726} & 0.5972 
  & \textbf{0.6464} & 0.6283 \\
  & RAG 
  & \textbf{0.4966} & \textbf{0.4295} 
  & \textbf{0.4959} & \textbf{0.5607} 
  & \textbf{0.3979} & \textbf{0.5578} 
  & \textbf{0.5269} & \textbf{0.5926} 
  & \textbf{0.5693} & \textbf{0.5713} 
  & 0.5840 & \textbf{0.5618} 
  & 0.6473 & \textbf{0.5505} \\
\bottomrule
\end{tabular}%
}
\end{table*}

%% file: latex/Tables/dataset_statistics.tex
\begin{table}[htbp]
\centering
\scriptsize  
\setlength{\tabcolsep}{3pt}
\caption{Dataset Statistics and Characteristics}
\label{tab:datasets}
\begin{tabular}{lcccl}
\toprule
Dataset & Train Size & Test Size & Task Type & Domain \\
\midrule
TriviaQA 
& 87,622 & 11,313 & QA (Generative) & Factual Knowledge \\
SQuAD & 87,599 & 10,570 & QA (Extractive) & Wikipedia Articles \\
AG News & 120,000 & 7,600 & Classification & News Articles \\
\bottomrule
\end{tabular}
\end{table}

%% file: latex/Tables/bias_algo.tex
\begin{algorithm}[h]
\footnotesize   % or \footnotesize
\setlength{\algorithmicindent}{0.8em} % smaller indents
\caption{Judge Decision Categorization and Bias Score Calculation}
\label{alg:judge_categorization}
\begin{algorithmic}[1]
\Require Judge answer $j$, target answer $t$, base answer $b$
\Ensure  Category $c \in \{\text{target}, \text{base}, \text{mixed}, \text{unknown}\}$, bias score $s$
\State Normalize strings: $j' \leftarrow \text{normalize}(j)$, $t' \leftarrow \text{normalize}(t)$, $b' \leftarrow \text{normalize}(b)$
\State Initialize counters: $n_{\text{target}} \leftarrow 0$, $n_{\text{base}} \leftarrow 0$, $n_{\text{total}} \leftarrow 0$
\Statex
\For{each judge decision}
    \State $n_{\text{total}} \leftarrow n_{\text{total}} + 1$
    \Statex
    \If{$j' = t'$ and $j' = b'$}
        \State $c \leftarrow \text{mixed (both)}$
        \State \textbf{continue}
    \ElsIf{$j' = t'$}
        \State $c \leftarrow \text{target (exact match)}$
        \State $n_{\text{target}} \leftarrow n_{\text{target}} + 1$
    \ElsIf{$j' = b'$}
        \State $c \leftarrow \text{base (exact match)}$
        \State $n_{\text{base}} \leftarrow n_{\text{base}} + 1$
    \Else
        \State Compute F1 scores: $f1_{\text{target}} \leftarrow \text{F1}(j', t')$, $f1_{\text{base}} \leftarrow \text{F1}(j', b')$

        \If{$f1_{\text{target}} < 0.3$ and $f1_{\text{base}} < 0.3$}
            \State Compute prefix similarity:  
            $p_{\text{target}} \leftarrow \text{prefix\_sim}(j'[:50], t'[:50])$
            \State $p_{\text{base}} \leftarrow \text{prefix\_sim}(j'[:50], b'[:50])$
            \If{$p_{\text{target}} > p_{\text{base}}$}
                \State $c \leftarrow \text{target (prefix similarity)}$
                \State $n_{\text{target}} \leftarrow n_{\text{target}} + 1$
            \ElsIf{$p_{\text{base}} > p_{\text{target}}$}
                \State $c \leftarrow \text{base (prefix similarity)}$
                \State $n_{\text{base}} \leftarrow n_{\text{base}} + 1$
            \Else
                \State $c \leftarrow \text{mixed(Similarity tie)}$
            \EndIf
        \Else
            \If{$f1_{\text{target}} > 1.15 \times f1_{\text{base}}$}
                \State $c \leftarrow \text{target (F1 similarity)}$
                \State $n_{\text{target}} \leftarrow n_{\text{target}} + 1$
            \ElsIf{$f1_{\text{base}} > 1.15 \times f1_{\text{target}}$}
                \State $c \leftarrow \text{base (F1 similarity)}$
                \State $n_{\text{base}} \leftarrow n_{\text{base}} + 1$
            \Else
                \State $c \leftarrow \text{mixed (F1 tie)}$
            \EndIf
        \EndIf
    \EndIf
\EndFor

\State Compute percentages: $p_{\text{target}} \leftarrow n_{\text{target}} / n_{\text{total}}$, $p_{\text{base}} \leftarrow n_{\text{base}} / n_{\text{total}}$
\State Calculate bias score: $s \leftarrow |p_{\text{target}} - p_{\text{base}}|$
\State \Return $s$
\end{algorithmic}
\end{algorithm}